\documentclass{article}

\usepackage{arxiv}

\usepackage[utf8]{inputenc} 
\usepackage[T1]{fontenc}    
\usepackage{hyperref}       
\usepackage{url}            
\usepackage{booktabs}       
\usepackage{amsfonts}       
\usepackage{nicefrac}       
\usepackage{microtype}      
\usepackage{lipsum}		
\usepackage{graphicx}
\usepackage{natbib}
\usepackage{doi}

\title{A ChatGPT-based approach for questions generation in higher education}

\date{June 10, 2024}	

\author{ \href{https://orcid.org/0009-0004-8161-1686}{\includegraphics[scale=0.06]{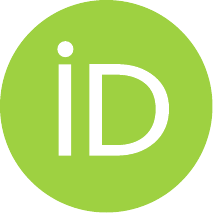}\hspace{1mm}Sinh Trong Vu} \\
	Banking Academy of Vietnam\\
	Hanoi, Vietnam \\
	\texttt{sinhvt@hvnh.edu.vn} \\
	\And
	\href{https://orcid.org/0000-0000-0000-0000}{\includegraphics[scale=0.06]{orcid.pdf}\hspace{1mm}Huong Thu Truong} \\
	Banking Academy of Vietnam\\
	Hanoi, Vietnam\\
	\texttt{trgthuhg@gmail.com} \\
	\AND
	Oanh Tien Do \\
	Banking Academy of Vietnam\\
	Hanoi, Vietnam \\
	\texttt{oanhvy2bg@gmail.com} \\
	\And
	Tu Anh Le \\
	Banking Academy of Vietnam\\
	Hanoi, Vietnam\\
	\texttt{leanhtu.work@gmail.com} \\
	\And
	Tai Tan Mai \\
	Dublin City University \\
	Dublin, Ireland \\
	\texttt{tai.tanmai@dcu.ie} \\
}

\renewcommand{\shorttitle}{\textit{arXiv} Template}

\hypersetup{
pdftitle={A template for the arxiv style},
pdfsubject={q-bio.NC, q-bio.QM},
pdfauthor={Sinh Vu Trong, Tai Tan Mai},
pdfkeywords={First keyword, Second keyword, More},
}

\begin{document}
\maketitle

\begin{abstract}
	Large language models have been widely applied in many aspects of real life, bringing significant efficiency to businesses and offering distinctive user experiences. In this paper, we focus on exploring the application of ChatGPT, a chatbot based on a large language model, to support higher educator in generating quiz questions and assessing learners. Specifically, we explore interactive prompting patterns to design an optimal AI-powered question bank creation process. The generated questions are evaluated through a "Blind test" survey sent to various stakeholders including lecturers and learners. Initial results at the Banking Academy of Vietnam are relatively promising, suggesting a potential direction to streamline the time and effort involved in assessing learners at higher education institutes.

\end{abstract}

\keywords{Large language model \and ChatGPT \and question generation}

\section{Introduction}
\label{sec:intro}
One of the most prominent advances in Artificial Intelligence recently is the development of large language models (LLMs), such as ChatGPT, BingChat, and Bard (developed by OpenAI, Microsoft, and Google, respectively). It can be said that large language models have been developing strongly in the past 2 years and creating a strong influence in the field of Generative AI (GenAI) \cite{Jahic2023}.

LLMs are capable of solving a wide range of tasks, such as natural language understanding, text generation, and sentiment analysis in various domains \cite{chang2023survey}. Since the breakthroughs of AI and LLM, education, as a crucial role in shaping society through almost every single individual, might receive significant benefits from these new LLM-based initiatives. AI is transforming education, bridging its gaps, and promoting a more inclusive and productive learning environment by customizing learning experiences, automating administrative tasks, and providing real-time feedback \cite{Kamalov}. 

Advancements in AI and LLMs have fueled the development of many educational technology innovations that aim to automate the often time-consuming and laborious tasks of generating and analyzing textual content, including generating test questions. For the application of LLM in education, this technology holds great potential in this field at all levels, especially university level \cite{Yan2023}. Therefore, this study conducts a research to comprehensively understand the capabilities of LLMs, especially ChatGPT when supporting lecturer to build question banks for learning modules in university setting. During the literature review, we witnessed that although the LLMs can provide quite a lot of support in teaching activities \cite{Daoetal2023}, the work of testing and evaluating learners, which takes up a lot of time and effort of lecturers, has not been studied in-depth by previous works. Therefore, in this paper, we decide to conduct research on the application of LLMs for generating question banks, with initial results at the Banking Academy of Vietnam. Specifically, the contributions of the paper are summarized as follows:

\begin{enumerate}
    \item Investigate the most potential LLM-based tool that satisfy the following criteria: (i): capable of processing multiple languages; (ii) widely accessible by the public and (iii) suitable for university educational setting.
    \item Design a novel approach to generate \textit{prompting patterns} to interact with ChatGPT, thereby design an effective prompt pattern for creating question banks at university level.
    \item Generate a question bank for a specific subject with diverse question types as the sample module of university level in Vietnam, in order to present the potential direction for further related research in the field of broader modules of higher education.
    \item Design and conduct a “\textit{Blind Test}” survey aimed at students to evaluate the quality of the generated questions.
\end{enumerate}

\section{Related works}
\label{sec:related}
With the growing interest in LLM for education, many studies around the world focusing on this issue have been conducted and have produced some interesting results. A group of authors from the University of Minnesota Law School have conducted research on ChatGPT's performance in answering a set of questions including 95 multiple-choice questions and 12 essay questions related to four undergraduate-level subjects in the law class. The results showed that ChatGPT studied and performed on the test at a level equivalent to a C+ grade for college students \cite{choi_hickman_monahan_schwarcz_2023}. 

Another study put ChatGPT to the test in completing the Dutch secondary school exam in the subject of English reading comprehension. This study concluded that ChatGPT achieved results with an average score of 7.3, comparable to the average score of all high school students in the country \cite{winter_2023}. 

In another research on evaluating the performance of LLMs when taking a high school level Biology test by the author namely Dao Xuan Quy, ChatGPT showed positive results with an accuracy of 71\% for remembering level questions and 61.82\% for understanding level questions. This result shows how effective ChatGPT is in capturing and clarifying concepts related to the required subject \cite{Dao2023}. 

The use of LLMs in supporting the construction of questions to test and evaluate students' knowledge should be encouraged. While exam-style questions are an important instructional tool for several reasons, manually creating questions is a time-consuming process that necessitates expertise, experience, and resources. This, in turn, impedes and inhibits the implementation of instructional activities (e.g., offering practice questions) and educational initiatives (e.g., adaptive testing) that need a huge pool of questions. Therefore, automatic question generating (AQG) approaches based on AI were established for research in both developed and developing nations \cite{kurdi2019}.

The general state of scientific research on the use of LLM in education is limited because its implementation in educational settings is still in its early stages. \textit{To the best of our knowledge, there is no study on evaluating quality and praticality of using questions generated by LLMs to test real students, especially in higher education}. Therefore, in this study we will conduct an experiment on questions bank generation using LLMs, with some early results from a university in Vietnam.

\textit{\textbf{Selecting the optimal LLM-based solution }}. 
Nowadays, many advanced large language models can be widely applied in the field of education such as ChatGPT, BingAI, Gemini, etc. Most prominently, ChatGPT is an advanced general AI model based on a Generative Pretrained Transformer (GPT). This feature refers to the LLM being pre-trained with the available dataset and can generategrammatically and contextually human-like responses according to the user's request through natural language processing. The current most popular versions of ChatGPT are GPT-3.5 and GPT-4, with the upgraded version GPT-4 can be used with a fee, which according to OpenAI performs much better with a 10 times larger pre-training datasets capacity \cite{KALYAN2024100048}.

The initial goal of this research is to evaluate various LLMs in order to find the one having performance and suitability for generating questions in the higher educational context. Each model has its own advantages and disadvantages, so our team has to compare consider which one can fulfill the both factors. As we prioritize the best ones meeting educational standards, ChatGPT Plus and Claude are two potential options. However, ChatGPT Plus has a relatively high fee (\textit{\$20/month}) and Claude has a few limitations in its capabilities \cite{Lozi2023}.

For the following reasons, we finally decided to choose ChatGPT 3.5 as the model for this research:
\begin{itemize}
    \item ChatGPT 3.5 is freely usable so this choice can save on our research costs.
    \item ChatGPT 3.5 has been proven through scientific research to be able to produce high-quality text, similar to text written by humans\cite{RAY2023121}.
    \item ChatGPT 3.5 currently has 180 million users, with 100 million weekly active users and 1.6 billion website visit times \cite{meer2024}. This number shows that ChatGPT 3.5 is sufficiently popular and trustworthy.

\end{itemize}
Striking the perfect balance between capability and cost, \textit{ChatGPT 3.5 is selected} as it is sufficiently suitable to the context of educating in the Banking Academy of Vietnam.

\section{Research design}
\label{sec:research}

\subsection{Choosing a subject for experimentation}
In most universities, building a question bank is the primary way to facilitate the procedure of making tests and final semester exams. The question bank is usually based on the learning outcomes and lesson content, thereby determining the types of questions to create. Questions are arranged according to each chapter, with cognitive levels depending on the objectives and lesson knowledge. Here, we conduct research on creating questions for the subject \textbf{\textit{"Corporate Finance I"}} with the LLM ChatGPT 3.5, by these reasons and purposes:
\begin{itemize}
    \item First, to test and evaluate ChatGPT 3.5's natural language processing ability with a variety of different question types. This is because “Corporate Finance I” has a variety of question types that have been extensively put under tests and final exams: Multiple Choice Questions (MCQs), True-False statements, Real-Scenario/Calculative exercises,...
    \item Second, to make the experiment more effective with the highest possible number of survey participants. “Corporate Finance I” is a popular module format as it is a compulsory curriculum for almost every faculty in the Banking Academy of Vietnam for specialized learning outcomes towards the students of the academy.
    \item Third, to qualifiedly evaluate this scientific study. Throughout many subjects our research team discussed, “Corporate Finance I” has been chosen since this is the subject that we have obtained a certain basic understanding of.
\end{itemize}
To conduct research, we combine personal knowledge and use documents including the textbook “Corporate Finance I” (Author \& Editor: Le Thi Xuan), the workbook “Corporate Finance I” of the Banking Academy and also refer to Course Learning Outcomes (CLOs), syllabus structure, test question formats, and examples of exam questions in this subject.

\subsection{The investigation of appropriate Prompt patterns}
Through research on the meaning of elements as well as the operation of available Prompt templates such as RTF (Role, Task, Format), RISE (Role, Input, Steps, Expectation), RTCF (Role, Task, Context, Format), RTCEF (Role, Task, Context, Example, Format). We have filtered, selected and separated into specific factors as shown in the Table \ref{tab:tab_prompt_elements} \cite{Cavoj2023}.

\begin{table}[h]
    \centering
    \caption{ Prompt elements and their uses}
  \begin{tabular}{|c|c|}
    \toprule
    \textbf{Factor}  & \textbf{Purpose}\\
    \midrule
    \textbf{Task}    & \begin{tabular}[c]{@{}c@{}}Describe the task you want the AI\\  to perform, usually starting with verbs \\ such as: Create {[}…{]}, do {[}…{]}, … This is \\ a required part in most prompts to \\ use for ChatGPT\end{tabular} \\
    \hline
    \textbf{Context} & \begin{tabular}[c]{@{}c@{}}Provide information about the context \\ and task objectives\end{tabular}\\
    \hline
    \textbf{Input}   & Describe the context in more detail\\
    \hline
    \textbf{Format}  & \begin{tabular}[c]{@{}c@{}}Specify the desired format for\\ the output, usually a standard that has been\\  agreed upon and used\end{tabular}\\
    \hline
    \textbf{Example} & \begin{tabular}[c]{@{}c@{}}Provide specific example instructions \\ to guide the AI, often in parentheses\end{tabular} \\
    \hline
    \textbf{Role}    & \begin{tabular}[c]{@{}c@{}}Clearly define the role of AI in \\ a specific context, usually with role \\ statements such as: Play the role of {[}…{]}, \\ you are {[}…{]}, …\end{tabular}\\
    \bottomrule
    \end{tabular}
    \label{tab:tab_prompt_elements}
\end{table}
Then, one by one, we select and combine the above factors and then test and evaluate the results returned by ChatGPT ourselves. The goal of this experiment is to answer: Whether to continue adding or removing any factor to see if it affects the quality of returned results? If the results are better, we consider that factor to be kept and conversely, if there is no change and we find it unnecessary, we remove that factor. As a result, we choose the most suitable prompt type for each type of question we want to create below.

\subsubsection{For multiple-choice questions}
\textit{Prompt: Role + Task + Context + Example + Format
}

Following the structure, here is our formative prompt:

\textit{Role}: You are a lecturer of [subject name]

\textit{Task}: Please create [a specific number] questions

\textit{Context}: [The kind of exercise] questions that focus on the content of [name of the lesson to test]

\textit{Example}: (This is the example I want you to imitate: [specific example of the type of exercise to create])

\textit{Format}: Present output in [format name]

\subsubsection{For True-False statements}
\textit{Prompt:= Role + Task + Context + Example + Format
}

Following the structure, here is our formative prompt:

\textit{Role}: You are a lecturer of [subject name]

\textit{Task}: Please create [a specific number] questions

\textit{Context}: [The kind of exercise]. You should make the exam of [number of statements] including [number of correct statements] correct statements and [number of incorrect statements] incorrect statements with content about [name of content]

\textit{Example}: (This is the example I want you to imitate: [specific example of the type of exercise to create])

\textit{Format}: All comments are in the same paragraph. The line determining the truth/falseness and explanation of that statement must immediately follow, starting with “ANSWER: ” (NOTE: Space after the colon) and then giving the appropriate answer and explanation.

\subsubsection{For Real-Scenario/Calculative exercises}
\textit{Prompt:= Task + Context + Input + Tone
}

Following the structure, here is our formative prompt:

\textit{Task}: Create exercises with specific data

\textit{Context}: About [name of content]

\textit{Input}: The topic includes [starting data] [detailed information]

\textit{Tone}: You have to solve the exercise as you are dealing with real business context

\subsection{Evaluation methods and test results}
\subsection{Preliminary self-assessment and optimization utilizing ChatGPT.}
Applying the above prompt samples with ChatGPT3.5, we obtain a list of questions, in which the MCQ and True/False questions are along with the solutions. We perform a preliminary self-assessment by checking whether the question list containing any duplication or lack of assumption to solve. From a total of 390 questions generated, we found and remove 74 ones with this strategy, in which, 28\% of the calculation exercises are insufficient data. For instance, the exercise \textit{"PQR Company wants to optimize ordering costs for product D. The annual demand is 600 tons, the purchase price per ton is 15,000 USD. Holding cost is 1.5\% of inventory value/year. Calculate the optimal order level."} lacks the \textit{"ordering costs"} value, makes it unsolvable. This is one of the drawback of large language model based method, relying on the probability of word's appearance rather than the logical factors. We show the self-assessment result in Table \ref{tab:evaluation_question} below.

\begin{table}[h]
    \centering
    \caption{Evaluation of question quality. (Unit: \%)}
   \begin{tabular}{|c|c|c|c|}
\hline
\textbf{Question Type} & \textbf{MCQs} & \textbf{\begin{tabular}[c]{@{}c@{}}True-False\\ Statement\end{tabular}} & \textbf{\begin{tabular}[c]{@{}c@{}}Calculation\\ exercise\end{tabular}} \\
\hline
\textbf{\begin{tabular}[c]{@{}c@{}}Total Number\\ of Questions\end{tabular}} & \begin{tabular}[c]{@{}c@{}}230\\ questions\end{tabular} & \begin{tabular}[c]{@{}c@{}}100\\ questions\end{tabular}& \begin{tabular}[c]{@{}c@{}}60\\ question\end{tabular}\\ 
\hline
\textbf{Duplicated} & 19,13\%  & 3\% & 0 \\
\hline
\textbf{\begin{tabular}[c]{@{}c@{}}Insufficient\\ assumptions\\ to solve\end{tabular}} & 3,48\% & 4\%  & 28\% \\
\hline
\end{tabular}
    \label{tab:evaluation_question}
\end{table}

\subsection{The "Blind Test" method}
To evaluate the quality of the questions more objectively, our research team conducted a survey using the “Blind test” method to collect answers from students to lecturers. Based on the well-known Turing Test, the "Blind test" aims to get empirical insight into people's ability to discriminate between artificial and human content \cite{KOBIS2021106553}.

\begin{figure}[h]
    \centering
    \includegraphics[width=0.5\textwidth]{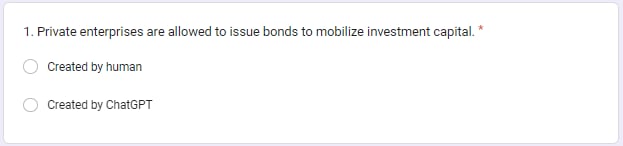}
    \caption{An example question from the “Blind Test” survey}
    \label{fig:eg_blind_test}
\end{figure}

Based on this idea, we have prepared two sets of question banks. One is the questions created by ChatGPT 3.5 and the other is the questions in the workbook of Corporate Finance 1 published by the Banking Academy of Vietnam. The questions from the two groups were mixed together. We proceed to create a survey form testing with a list of 15 questions, with an example shown in Figure \ref{fig:eg_blind_test}. Each question includes 2 options: created by humans or created by ChatGPT. We obtain the answer from a total of 91 people including lecturers, students who studied, are currently studying or have not studied Corporate Finance I before. The detail statistics about this survey is evaluated in the following sections.

\subsection{Result of evaluation}
\subsubsection{Survey assumption}

During the survey, the course-completed rate is expected to be distributed in proportion: 60\% have completed and are in progress of completing, and 40\% have not started. This provides the opportunity for multi-pronged analysis looking at different outcomes based on these two main groups.

The general expectation of our research team is that the number of questions prepared by ChatGPT but predicted to be created by humans is high, for approximately over 60\%. To better understand this result, we need to analyze each group more carefully as follows:
\begin{itemize}
    \item For lecturers, they have the ability to deeply evaluate the similarity between the question and the content, evaluate the difficulty level of the question as well as the feasibility of solving it. This allows them to most accurately identify which questions truly reflect knowledge and are capable of assessing students' skills like real exam questions.
    \item For the group of people who have studied, they can analyze the content of the questions, consider the level of difficulty and the ability to apply learned knowledge to solve questions. This helps them better recognize how each knowledge is asked in the real exam, thereby pointing out the questions created by ChatGPT with its shortcomings.
    \item The group of in-progress learners can focus on evaluating the reflectiveness of the question on the content they are approaching. From their perspective, they will distinguish between questions created by ChatGPT or people based on their progressed knowledge and their own understanding.
    \item Finally, the group of students that haven’t learned this subject may only be able to evaluate the naturalness and the grammar accuracy in the question.
\end{itemize}
\subsubsection{General analysis}
\begin{figure}[h]
    \centering
    \includegraphics[width=0.45\textwidth]{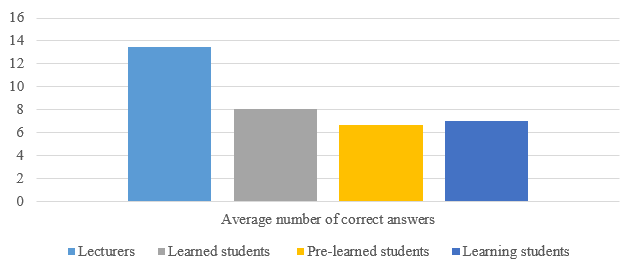}
    \caption{Average number of correct answers from different groups of surveyors}
    \label{fig:average_answer}
\end{figure}

The performance of each group of survey participants is presented in the Figure \ref{fig:average_answer}.

\begin{itemize}
    \item The survey results showed that lecturers have a relatively high ability to distinguish between questions generated by ChatGPT and those generated by humans, with an accuracy rate of up to 13.5/15 questions.
    \item Students who have studied the subject are better at distinguishing between ChatGPT-generated and human-generated questions than students who are currently studying (8.0833 > 7), and students who have studied have a higher average correct answer rate than students who have not studied (7 > 6.7059). This suggests that exposure to course content and knowledge helps students become more sensitive to the differences between human-generated and AI-generated questions.
\end{itemize}

\subsubsection{Specific analysis}

Our data shows that the 9 questions having the least percentage of students answering correctly have an average of only 32.5\% answers correct. In particular, the survey showed that the multiple-choice section had the lowest rate of correct answers with only about 30\%. This may stem from the fact that human lecturers often create short, concise multiple-choice questions with the goal of simply testing whether students have grasped the knowledge surrounding the concept being asked. In fact, LLM like ChatGPT has the ability to grasp the subject and create related questions to support learners\cite{Dao2023}. Therefore, this tool is capable of creating MCQs properly along with the lecturers in an even faster way.

\begin{table}[h]
    \centering
    \caption{The best questions that ChatGPT generates}
   \begin{tabular}{|l|l|}
\hline
\begin{tabular}[c]{@{}l@{}}1. The principle of time \\ value of money is:\end{tabular}& \begin{tabular}[c]{@{}l@{}}2. What are some of \\ the main benefits of \\ long-term investing?\end{tabular}\\
\begin{tabular}[c]{@{}l@{}}A. The longer a currency \\ retains its value, the better\end{tabular} & \begin{tabular}[c]{@{}l@{}}A. High short-term \\ profits.\end{tabular}\\
\begin{tabular}[c]{@{}l@{}}B. The value of money\\ increases over time\end{tabular} & \begin{tabular}[c]{@{}l@{}}B. Increase the value\\  of long-term assets.\end{tabular}\\
\begin{tabular}[c]{@{}l@{}}C. A currency today has\\  a higher value than it \\ will in the future\end{tabular} & C. Low risk.\\
\begin{tabular}[c]{@{}l@{}}D. Money loses value \\ over time\end{tabular}& D. Stable market.\\ 
\hline
\begin{tabular}[c]{@{}l@{}}Participant result: \\ 63 (69,2\%) wrongly \\ rated as Human, 28 \\ (30,8\%) correctly rated \\ as ChatGPT\end{tabular} & \begin{tabular}[c]{@{}l@{}}Participant result:\\  67 (73,6\%) wrongly \\ rated as Human, 24\\ (26,4\%) correctly rated \\ as ChatGPT\end{tabular} \\ \hline
\end{tabular}
    \label{tab:best_question}
\end{table}
To explain the results above, the differentiating difficulty of AI-generated exercise questions is based on two factors:
\begin{itemize}
    \item The question type: Whether it is an MCQ or a T/F, Exercise question.
    \item The quality requirement of information input: In the process of making a good question, whether the question is required to capture completely learning materials or a few texts for example.
\end{itemize}
Subjectively stated, a broad and general list of MCQs surveyed are less difficult to be guessed than real-life scenario exercises, or calculative ones that we had to give ChatGPT some examples to train it. Furthermore, these models can generate plausible distractors, which are incorrect answers that are designed to be similar to the correct answer, making it more challenging for humans to discern the difference. According to the study of Bitew et al. (2023), ChatGPT, when guided by question items and in-context examples, can generate high-quality distractors that are suitable for immediate use in an educational context.\cite{bitew2023distractor}

\begin{table}[h]
    \centering
    \caption{Worst question ChatGPT creates}
   \begin{tabular}{|l|l|}
\hline
{\begin{tabular}[c]{@{}l@{}}Private enterprises are\\  allowed to issue bonds \\ to mobilize investment \\ capital. (True/False)\end{tabular}} & {\begin{tabular}[c]{@{}l@{}}Strengthening overdue \\ debt management helps \\ businesses optimize capital\\  resources. (True/False)\end{tabular}} \\ 
\hline
\begin{tabular}[c]{@{}l@{}}Participant result: 70 (76,9\%) \\ correctly rated as Human, \\ 21 (23,1\%) wrongly rated \\ as ChatGPT\end{tabular} & \begin{tabular}[c]{@{}l@{}}Participant result: 70 (76,9\%) \\ correctly rated as Human,\\  21 (23,1\%) wrongly rated\\  as ChatGPT\end{tabular} \\ 
\hline
\end{tabular}
    \label{tab:worst_question}
\end{table}
However, there are a few limitations needed to be considered to the performance of LLMs in generating questions in order to decide the optimal usage of this tool. Firstly, the 9 AI-generated questions, especially the True-False questions and Exercise questions that are leastly mistaken for human-generated ones shown, are all lacking structured links to other concepts and overall reasoning around the subject. This proves that the discriminatory power of those 9 questions might not be as high as other questions in the list. To clarify this point, our research team have tried to answer the questions ourselves as students and stated that more than half of the questions can be referenced for posting on the real exam, but there are a few problems: 
\begin{itemize}
    \item In regards of the questions’ content, they are simply the same so it is hard to use these questions to encourage students to link back what they have learned.
    \item Some of the AI-generated questions are quite vague to make clear its meaning.
    \item In the Vietnamese version of the questions, the way ChatGPT used related terms is not really exact for academic context due to the language barrier.
\end{itemize}
\section{Conclusion}
In this study, we have researched an overview of the major language models available today, through which we propose to use Chat GPT3.5 because it has a wide reach to users, suitable for the Practical events at the Banking Academy. We have researched and extracted effective command samples that interact with Chat GPT3.5 to create a question bank for Corporate Finance 1 subject at the Banking Academy today and then create a bank. Questions types include: multiple choice, true or false judgment and calculation exercises. The question bank created by Chat GPT3.5 is post processed, then evaluated by a "Blind Test" conducted on  lecturers and students who studied, studying or have not studied this subject. The results show that ChatGPT has great potential in supporting learner assessment through the application of creating questions that contribute to the test bank. This conclusion is proven by accurate data, with a level of differentiation of about 80\% at an average level similar to the target of test banks created by instructors.


 During the research and implementation process, we encountered a limitation that is needed for other works. Currently, with the rapid development of artificial intelligence, many new tools have emerged to support users in creating question and answer systems for their specific purposes, which can upload user's documents. These documents play as a knowledge base which contributes to improving the quality of questions generated. Additionally, users only need to provide the initial prompt, and educators do not need to learn about prompts but can simply request questions based on quantity and topic. Therefore, we hope to integrate our methodology with other scientific solutions to save time and effort for lecturers in evaluating learners in higher education.

\bibliographystyle{plain}
\bibliography{references}  






\end{document}